\documentclass[usenatbib,useAMS]{mn2e}
\usepackage{times}
\usepackage{amsmath}
\usepackage{amssymb,amsfonts}
\usepackage{graphicx}
\usepackage{verbatim}
\usepackage{psfig}
\usepackage{epsf}
\newcommand{\ltaraw}{$\; \buildrel < \over \sim \;$}
\newcommand{\lta}{\lower.5ex\hbox{\ltaraw}}
\newcommand{\gtaraw}{$\; \buildrel > \over \sim \;$}
\newcommand{\gta}{\lower.5ex\hbox{\gtaraw}}

\newcommand{\qso}{Q2237+0305}

\loadboldmathitalic 
\title [Microlensing of BLRs]
{Gravitational microlensing of quasar broad line
regions at large optical depths}
\author[G. F. Lewis \& R. A. Ibata]
{Geraint F. Lewis$^{1}$ \& R. A. Ibata$^{2}$\\
$^{1}$
Institute of Astronomy,
School of Physics, University of Sydney, NSW 2006, Australia:
Email {\tt gfl@physics.usyd.edu.au}\\
$^{2}$
Observatoire de Strabourg, 11, rue de l'Universit\'e, F-67000, Strasbourg,
France:
Email \tt{ibata@astro.u-strasbg.fr}
}

\date{\today}
\begin{document} 
\maketitle 
\begin{abstract}
Recent estimates  of the scale of  structures at the  heart of quasars
suggest that  the region responsible  for the broad line  emission are
smaller  than previously thought.   With this  revision of  scale, the
broad  line  region is  amenable  to  the  influence of  gravitational
microlensing. This study investigates the influence on microlensing at
high optical  depth on a  number of current  models of the  Broad Line
Region (BLR). It is found  that the BLR can be significantly magnified
by the action of microlensing, although the degree of magnification is
dependent   upon  spatial   and  kinematic   structure  of   the  BLR.
Furthermore,  while there  is a  correlation between  the microlensing
fluctuations of the continuum source and the BLR, there is substantial
scatter  about  this relation,  revealing  that broadband  photometric
monitoring is not necessarily a guide to microlensing of the BLR.  The
results  of this  study  demonstrate that  the  spatial and  kinematic
structure  within   the  BLR  may  be   determined  via  spectroscopic
monitoring of microlensed quasars.
\end{abstract}

\section{Introduction}\label{introduction}
Quasars  are amongst  the most  luminous sources  in the  universe. At
cosmological  distances,  their  relatively  small  size  ensures  the
regions responsible for producing the various spectral line components
remains effectively unresolved  with modern telescopes.  Gravitational
microlensing,  however, can significantly  magnify the  inner regions,
providing  clues to  the various  scales of  structure located  at the
heart of  quasars, giving some of  the best estimates of  the scale of
the         central          continuum         emitting         region
\citep[e.g.][]{1992ApJ...396L..65J,1999ApJ...519L..31Y,2000MNRAS.315...62W},
as well  as offering  the possibility of  probing the nature  of other
quasar               small               scale               structure
~\citep{2000PASP..112..320B,2002ApJ...577..615W}.

The degree  of microlensing magnification is dependent  upon the scale
size of  the source,  with smaller sources  being more  susceptible to
large  magnifications  \citep[e.g.][]{1991AJ....102..864W}.  While  the
continuum  emitting region  of a  quasar  is small  enough to  undergo
significant magnification,  the more extensive  line emitting regions,
specifically  the  BLR with  a  scale length  of  0.1-a  few pc,  were
considered  to  be  too  large to  suffer  substantial  magnification.
\citet{1988ApJ...335..593N} undertook a  study to determine the degree
of microlensing of various models  of the BLR, examining the influence
of a single  microlensing mass in front of  the emission region.  When
considering  microlensing in  multiply imaged  quasars,  however, many
stars are  expected to influence the  light beam of  a distant source,
and these  combine in  a very non-linear  fashion and the  single star
approximation  is   a  poor  one  \citep[e.g.][]{1990ApJ...352..407W}.
\citet{1990A&A...237...42S}  considered the microlensing  of a  BLR at
substantial   optical   depth.   These   studies   found  that   while
gravitational microlensing  did result in the modification  of the BLR
emission line  profiles, the overall  magnification of the  region was
small, typically less than 30\%.

These  microlensing studies  employed  estimates of  the  size of  the
quasar    BLR    based   upon    simple    ionization   models    [see
\citet{1979RvMP...51..715D}]. Reverberation mapping, however, provides
a more  direct measure of  the geometry of  the BLR and  early studies
suggested these simple ionization  models had over estimated the scale
of    the     BLR    by    roughly    an     order    of    magnitude
\citep[e.g.][]{1985ApJ...292..164P},  prompting   a  revision  of  BLR
physics   \citep{1989ApJ...347..640R}.    More  recent   reverberation
measurements have  refined the  size of the  BLRs in  active galaxies,
finding  it  to  be  $\sim10^{-4}$pc  in low  luminosity  AGN,  up  to
$\sim10^{-1}$pc in luminous quasars, with  the size of the BLR scaling
with the luminosity of the quasar, such that $R_{BLR} \propto L^{0.7}$
\citep{1999ApJ...526..579W,2000ApJ...533..631K}.   Furthermore,  these
results  demonstrate the  BLR possesses  a stratified  structure, with
high ionization lines  being an order of magnitude  smaller than lower
ionization        lines.        Following        this       discovery,
\citet{2002ApJ...576..640A}    reexamined   the   question    of   the
microlensing  of  the BLR  region  in  light  of this  revised  scale.
Undertaking an  analysis similar to  \citet{1988ApJ...335..593N}, they
considered the influence of a  single microlensing mass located in the
BLR,  finding that significant  modification of  the BLR  line profile
results.

\begin{figure}
\centerline{ \psfig{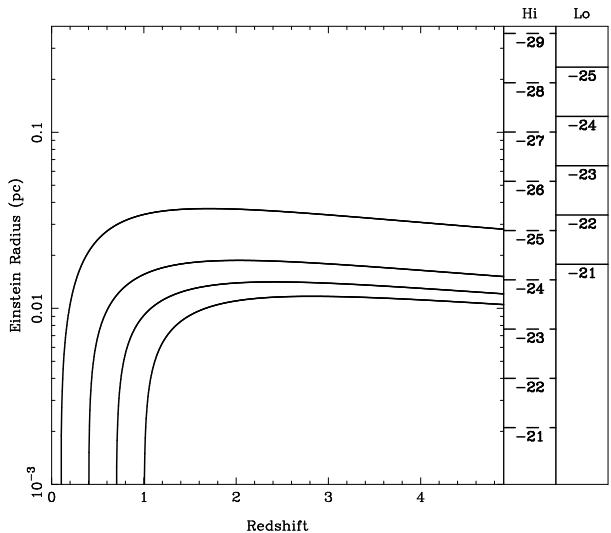}}
\caption[]{The Einstein  Radius for  a solar mass  star for  lenses at
$z=0.1,0.4,0.7$ \& $1.0$, denoted by the location of the start of each
curve, for  sources at a range  of redshifts. The panels  on the right
hand  size present  the expects  size of  the high  ionization (column
topped with Hi) and low ionization (Lo) BLRs, for the denoted absolute
magnitudes.
\label{fig1a}}
\end{figure}

As  with the  approach  of \citet{1988ApJ...335..593N},  the study  of
\citet{2002ApJ...576..640A} is  a poor representation  of microlensing
at  significant  optical  depth,  the situation  for  multiply  imaged
quasars.   This  paper,  therefore,  also  examines  the  question  of
microlensing   of   the   BLR,   extending  the   previous   work   of
\citet{2002ApJ...576..640A} into the higher optical depth regime.  The
approach  to  this  question  is  described  in  Section~\ref{method},
whereas  the  results  are  discussed  in  Section\ref{results}.   The
conclusions to this study are presented in Section~\ref{conclusions}.

\section{Method}\label{method}

\subsection{Microlensing Maps}\label{maps}
The known number of multiply  imaged quasars has been steadily growing
in recent years \citep{1998ApSS.263...51M}. For the majority of these,
little temporal data has been  obtained and so this study focuses upon
the   quadruple  quasar   Q2237+0305,  the   first  system   in  which
microlensing    was   confirmed    \citep{1989AJ.....98.1989I}.    The
microlensing  parameters, the  surface mass  density $\sigma$  and the
shearing due to  external mass $\gamma$, were taken  from the modeling
of  \citet{1998MNRAS.295..488S}; for  images C  and D  the  values for
$(\sigma,\gamma)$  were $(0.69,0.71)$ and  $(0.59,0.61)$ respectively.
Images  A and  B have  similar microlensing  parameters, chosen  to be
$(0.36,0.41)$  for  the purpose  of  this  study.  These  microlensing
parameters may  seem quite  specific and hence  not applicable  to the
population of lensed quasars  in general. However, for multiply imaged
systems the microlensing parameters  must be substantial and hence the
simulations in this paper can be taken as representative.

\begin{figure*}
\centerline{ \psfig{figure=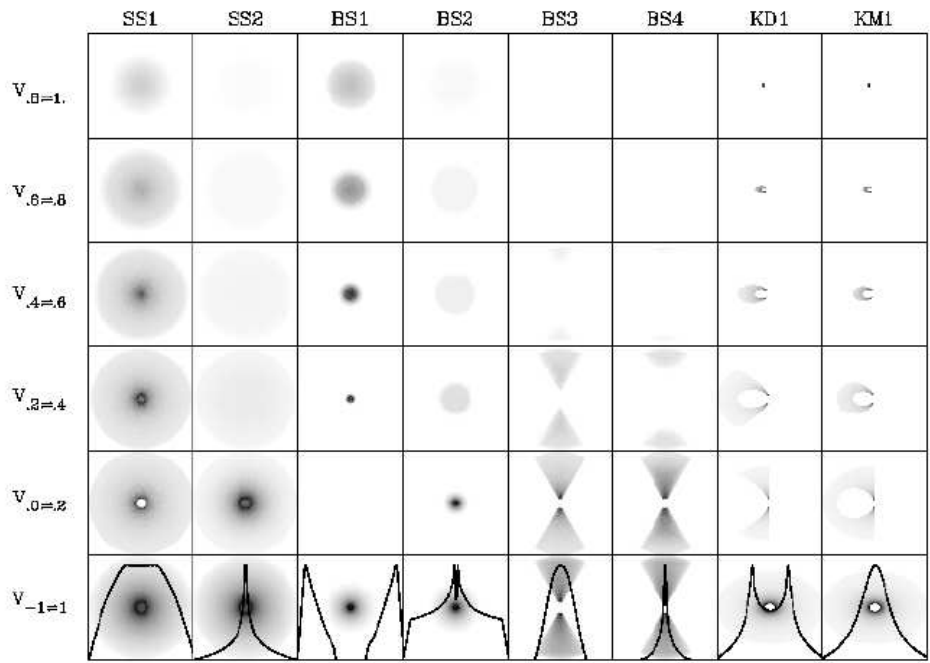,angle=270,width=5.8in}}
\caption[]{The appearance of  each of the BLR models  as a function of
velocity. The  models are presented  vertically, with the  model label
given at the  top of each column; the details of  the models are given
in  Table~\ref{table1},  with  the  SS models  representing  spherical
shells of clouds,  the BS models being biconical  shells, where as the
KD \& KM models are thin  disks. The five velocity slices consider the
range from  0 to 1  in units  of 0.2; as  the models are  symmetric in
velocity, the appearance of the BLR at negative velocities is the same
as positive  velocities (this is also  true of KD1 and  KM1 except the
images are  flipped about the  y-axis). The lowest panels  present the
velocity  integrated surface  brightness distribution  of each  of the
models. Superimposed upon these  are the spectral line profiles; these
can     be     compared    directly     with     the    results     of
\citet{2002ApJ...576..640A}.
\label{fig1}}
\end{figure*}

Furthermore, the  important length  scale in microlensing  problems is
the Einstein  radius (ER) of a  single star projected  onto the source
plane, with  sources substantially smaller than  this size susceptible
significant  magnification,  whereas larger  sources  are more  mildly
affected  by  lensing \citep{1991AJ....102..864W,1992ApJ...392..424W}.
For Q2237+0305, this length scale (for a solar mass star) is $\eta_o =
0.06   pc$\footnote{A  concordance  cosmology,   with  $\Omega_o$=0.3,
$\Lambda_o$=0.7 and $H_o=72$km/s/Mpc,  is assumed.}  and is comparable
to the revised scale of the BLR in quasars.  As Q2237+0305 remains one
of  the few  multiply imaged  quasars in  which microlensing  has been
unambiguously  detected,  it potentially  offers  the  best chance  of
observing the influence of microlensing of the BLR.

\citet{2002ApJ...576..640A}  presented a  detailed  discussion on  the
observability  of  BLR  microlensing  for various  gravitational  lens
systems. Figure~\ref{fig1a} presents the ER  for a solar mass star for
a  range  of  source  redshifts,  with lenses  at  $z=0.1,0.4,0.7$  \&
$1.0$. The two panels on the right hand side present the expected size
for the high ionization BLR (column topped with Hi) and low ionization
BLR  (Lo), for  a  range  of absolute  V-band  magnitudes; these  were
calculated  using  the  measured  BLR   sizes  of  NGC  5548  and  the
$R_{BLR}\propto                    L^{0.7}$                    scaling
\citep{1999ApJ...526..579W,2000ApJ...533..631K}.    Clearly,  the  low
ionization  BLR become  quite extensive  in moderately  bright quasars
($M_V \gta -25$), up to 10$\times$  larger than the ER for the majority
of cases. For some  configurations, with relatively nearby lenses, the
difference  is only  a  factor of  four,  a BLR  source  size that  is
investigated  in this  paper. The  size  of the  high ionization  BLR,
however,  is similar to  the ER  for many  lensing geometries,  and is
amenable  to microlensing for  quite luminous  quasars.  It  should be
noted  that the ER  scales with  the square  root of  the microlensing
mass;   even   accounting  for   a   typical   microlensing  mass   of
$\sim0.1M_\odot$,  the  high  ionization  BLR  for  relatively  bright
quasars  should  be  susceptible  to  microlensing.   As  an  example,
Q2237+0305,  the quasar  considered  in this  paper,  has an  absolute
magnitude of  $M\sim-26$ [accounting  for a magnification  of $\sim16$
\citep{1998MNRAS.295..488S}], with  its ER being  comparable the scale
of its high ionization BLR, although the low ionization BLR may be too
extensive to be significantly magnified.

\begin{figure*}
\centerline{ \psfig{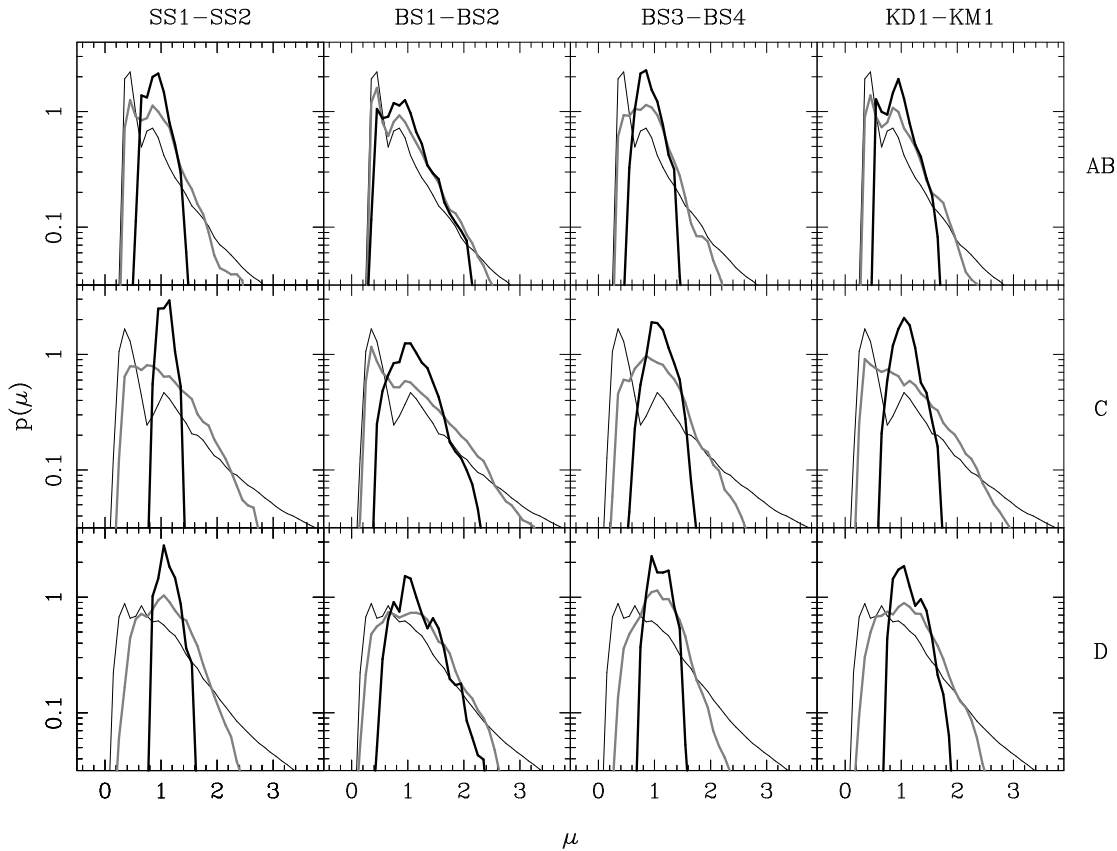}}
\caption[]{The magnification  distributions of the  various BLR models
of  each  of  the  quasar   images  in  Q2237+0305.   The  models  run
vertically, with the  model name at the top of  each column, while the
name of each quasar image is  presented down the right hand side.  The
thin line in each  panel corresponds to the magnification distribution
of a  point source.   The grey lines  correspond to  the magnification
distributions for  the BLR  models with a  radius of 1ER,  whereas the
black line is the magnification distribution for BLRs of radius 4ER.
\label{fig2}}
\end{figure*}

Magnification  maps  were constructed  using  a ray-tracing  algorithm
\citep{1986A&A...166...36K,1990ApJ...352..407W}.   For  each image,  a
square  region 20  ER on  a side  was generated.   The  resolution was
chosen such that  one ER corresponded to 128 pixels  and the number of
rays   traced  ensured  that   the  Poissonian   error  in   the  mean
magnification was less than 0.5\%.

\subsection{BLR Models}\label{blrmodel}
To address the  question of how microlensing influences  the BLR it is
important  to   determine  how   the  region  appears   at  particular
velocities.     To    undertake    this,    the    BLR    models    of
\citet{2002ApJ...576..640A}  were adopted;  the  mathematics of  these
models are presented in this  earlier paper and will not be reproduced
here.  The BLRs were constructed  via a Monte Carlo approach, randomly
distributing  large numbers  of clouds  with the  appropriate spatial,
emissivity  and velocity  characteristics. By  selecting  in velocity,
images of the source in  100 velocity slices were constructed.  As per
the earlier  study of  \citet{2002ApJ...576..640A}, two sizes  for the
BLR were considered,  a smaller source with a radius of  1 ER, as well
as  a larger source  of radius  4 ER.  A summary  of the  eight models
employed  in  this  study  is presented  in  Table~\ref{table1};  this
labeling will be used through the rest of this study.

\begin{table}
\centering
\begin{minipage}{70mm}
\caption{Summary of BLR models: For each model, p denotes the
radial form of the velocity field, such that $v \propto r^{\rm p}$,
while $\theta$ represents the orientation of non-spherically symmetric
models [see \citet{2002ApJ...576..640A}].\label{table1}}
\begin{tabular}{lc} \hline \hline
Name &  Description \\ \hline SS1  & Spherical Shell (p=0.5)  \\ SS2 &
Spherical Shell (p=2.0) \\ BS1 & Biconical Shell (p=0.5, $\theta=0^o$)
\\  BS2 &  Biconical Shell  (p=2.0, $\theta=0^o$)  \\ BS3  & Biconical
Shell  (p=0.5,  $\theta=90^o$)  \\   BS4  &  Biconical  Shell  (p=2.0,
$\theta=90^o$) \\ KD1 &  Keplerian Disk (p=-0.5, $\theta=45^o$) \\ KM1
& Modified Disk (p=-0.5, $\theta=45^o$) \\ \hline
\end{tabular}
\end{minipage}
\end{table}
\begin{figure*}
\centerline{ \psfig{figure=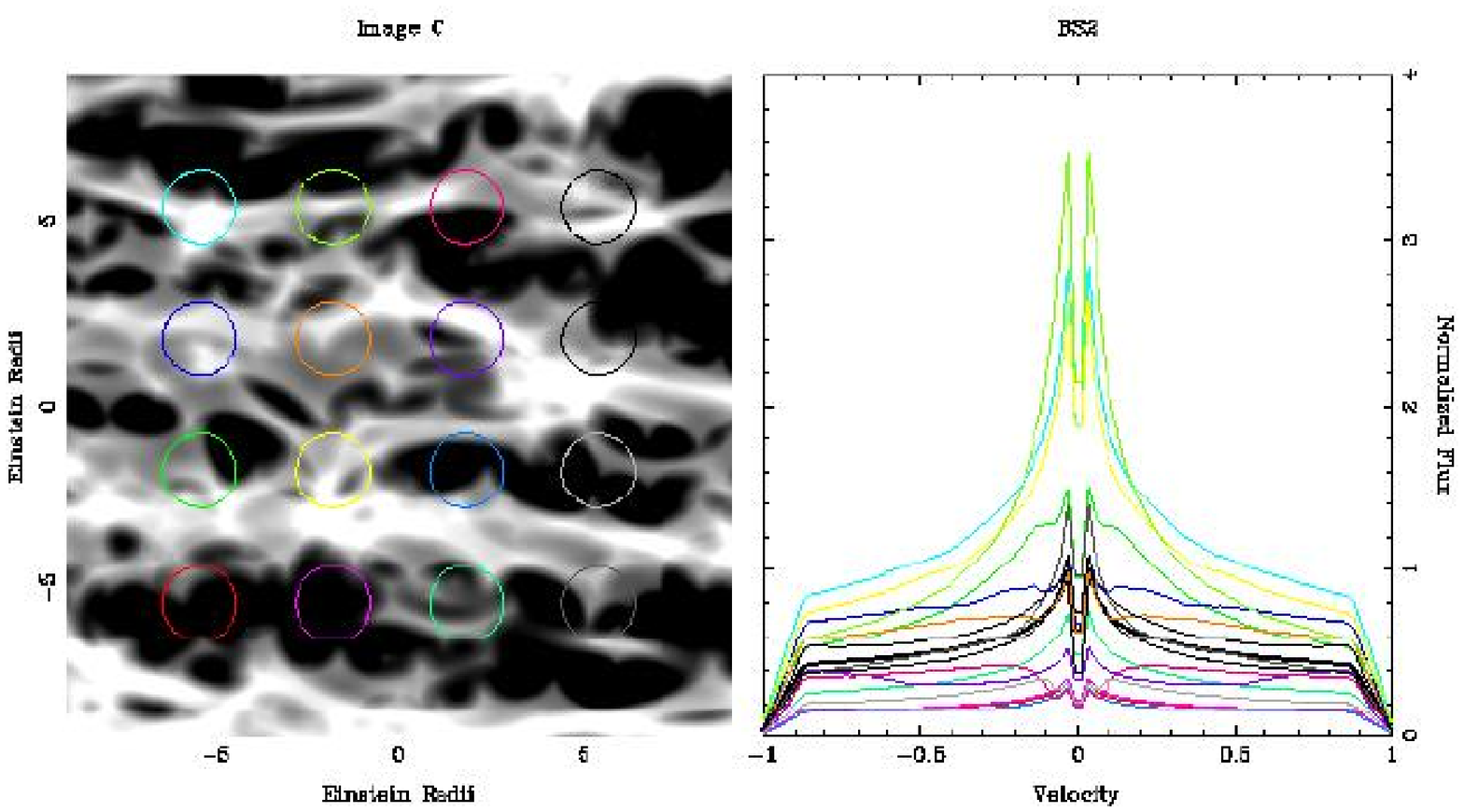,angle=270,width=5.8in}}
\caption[]{An example of the modification  of the BLR line profile for
the smaller BS2 for the microlensing model of Image C.  In this panel,
lighter  regions  denote  regions  of  magnification,  whereas  darker
regions correspond  to regions of  demagnification.  In the  left hand
panel, the  magnification map,  convolved with the  surface brightness
distribution of the  BLR, is presented. The right  hand panel presents
several microlensed  BLR line profiles for the  locations indicated by
the colour-coded circles over  the magnification map.  The thick black
line corresponds to the unmicrolensed  line profile. Note that, due to
the symmetric  nature of the model surface  brightness distribution in
velocity  space, the  microlensed  BLR line  profile  is symmetric  in
velocity.
\label{fig3}}
\end{figure*}

\section{Results}\label{results}
To   calculate  the  influence   of  the   gravitational  microlensing
magnification upon the BLR, each image of the source (as a function of
velocity) was  convolved with the  magnification maps. For  the larger
BLR models,  the source  size becomes comparable  to the scale  of the
magnification maps.   Hence, appropriate regions are  trimmed from the
convolved maps to  negate edge effects. For the  smaller sources, this
means that the  central 18$^2$ ER were employed  for analysis, whereas
the  larger sources yielded  a region  12$^2$ ER.   Note that  for the
purposes of this  study, all models are oriented  perpendicular to the
shear field.   Random orientations of  the BLR models with  respect to
the microlensing structure is reserved to a future contribution.

\subsection{Magnifications}\label{mags}
The magnification distributions  of the total flux of  the BLR models,
determined  by convolving the  velocity integrated  surface brightness
profile    with   the   magnification    maps,   are    presented   in
Figure~\ref{fig2}.  Note  that each panel  presents a pair  of models,
denoted at  the top of  each column; this  is because in each  pair of
models the  radial emission properties of  the clouds are  the same so
that the velocity integrated surface brightness profiles are the same.
This   can   be   seen   in   the   lowest   series   of   panels   in
Figure~\ref{fig1}. In each panel, three curves are given; the lightest
is  the magnification  distribution  of a  single  pixel, whereas  the
thicker, grey line is the distribution for the smaller BLR models. The
thick black line corresponds to the magnification distribution for the
larger BLR models.

As  the  source  size   increases,  the  width  of  the  magnification
probability  distribution  decreases; in  comparing  the single  pixel
source  with  the  smaller  BLR  model,  it is  clear  that  the  high
magnification   tail  has   been  curtailed~\footnote{Note   that  the
magnification distributions in this  paper are normalized with respect
to the  mean microlensing magnification  $\mu_{th} = (  (1-\sigma)^2 -
\gamma^2)^{-1}$; hence source  can undergo substantial demagnification
as well as  magnification and deviations are with  respect to the mean
behaviour of the emission line.}.  The smaller BLR model can, however,
suffer  significant magnification,  with the  total flux  in  the line
being boosted by a  factor of 1.5-2 in most of the  cases. On the face
of  it, this is  rather surprising  as the  source radius  of 1  ER is
relatively large.   It is important to remember,  however, that unlike
numerous  previous  microlensing  studies,  the  source  here  is  not
uniform, but possesses structure  on scales substantially smaller than
an ER and this can be more significantly magnified.

Examining the  magnification probability distributions  for the larger
BLR  models reveals  that  they too  can  be substantially  magnified,
although the  magnification distribution is narrower than  the case of
the smaller  BLR sources.  In most  cases, the BLR can  be enhanced by
$\sim50\%$.   Interestingly,  the   $BS1$-$BS2$  pair  of  models  are
particularly   broad   compared   to   the  other   cases;   examining
Figure~\ref{fig1}  reveals that  the surface  brightness distributions
for  these models  are quite  centrally concentrated  compared  to the
other  models, and  this small  scale structure  can  be substantially
magnified.   Again, this smaller  scale structure  of the  BLR surface
brightness  distribution  results  in  stronger magnification  than  a
uniform source of the same radius.

This is  further illustrated  in Figure~\ref{fig3} which  presents the
form of  the BLR emission  line profile for  the smaller BS2  model as
microlensed  by Image  C in  2237. The  left hand  panel  presents the
magnification   map  convolved   with  the   BS2   surface  brightness
distribution. The series of coloured circles over the map indicated 16
fiducial locations  over the map where  the form of  the emission line
profile were calculated. These are  presented in the right hand panel,
with the  solid black line  being the unlensed emission  line profile;
note that the  flux in the microlensed line  profiles has been divided
by the mean magnification of the microlensing map so that a meaningful
comparison between them and the unlensed case can be made.

\begin{figure*}
\centerline{ \psfig{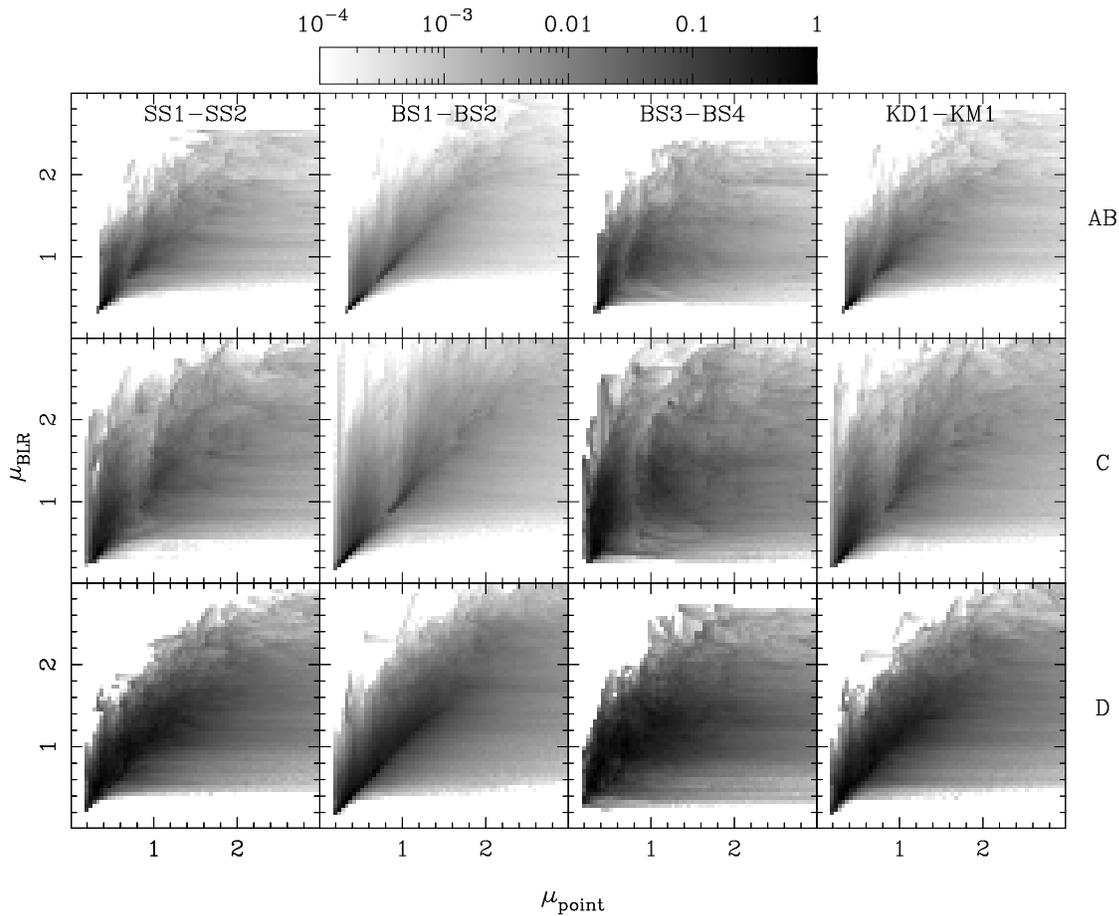}}
\caption[]{The correlation  between the magnification  of a point-like
source (here one  pixel) and the smaller BLR  models, for simultaneous
observations.   Note  that the  relative  probabilities are  displayed
logarithmically as denoted by the gray-scale at the top of the figure.
\label{fig4}}
\end{figure*}

It is clear that there is  substantial variation in not only the total
emission line flux, but also in the form of the emission line profile.
The  red source  in the  lower left  hand corner  lies primarily  in a
rather uniform  region of demagnification and while  its emission line
profile is clearly suppressed it  appears to have retained its overall
shape. The line  profile of the dark blue source on  the far left hand
side, while  being magnified in the  wings of the line,  appears to be
unmagnified at velocities near zero. Examining the source profile as a
function of  velocity (Figure~\ref{fig1}) it is apparent  that the BS2
model  is  very  centrally   concentrated  at  velocities  near  zero,
appearing more extensive at  the velocity extremes. When examining the
magnification map  in the vicinity of  the blue source it  can be seen
that  the  outer regions  overlay  strong  magnification, whereas  the
central regions lie in regions of mean magnification.  The opposite is
true for the  light green source at the top of  the second column from
the left. Here, the central regions  of the BLR lie within a region of
strong magnification, whereas the outer regions are are less affected,
leading  to a  strong  enhancement  of the  emission  line profile  at
velocities near  zero.

\begin{figure*}
\centerline{ \psfig{figure=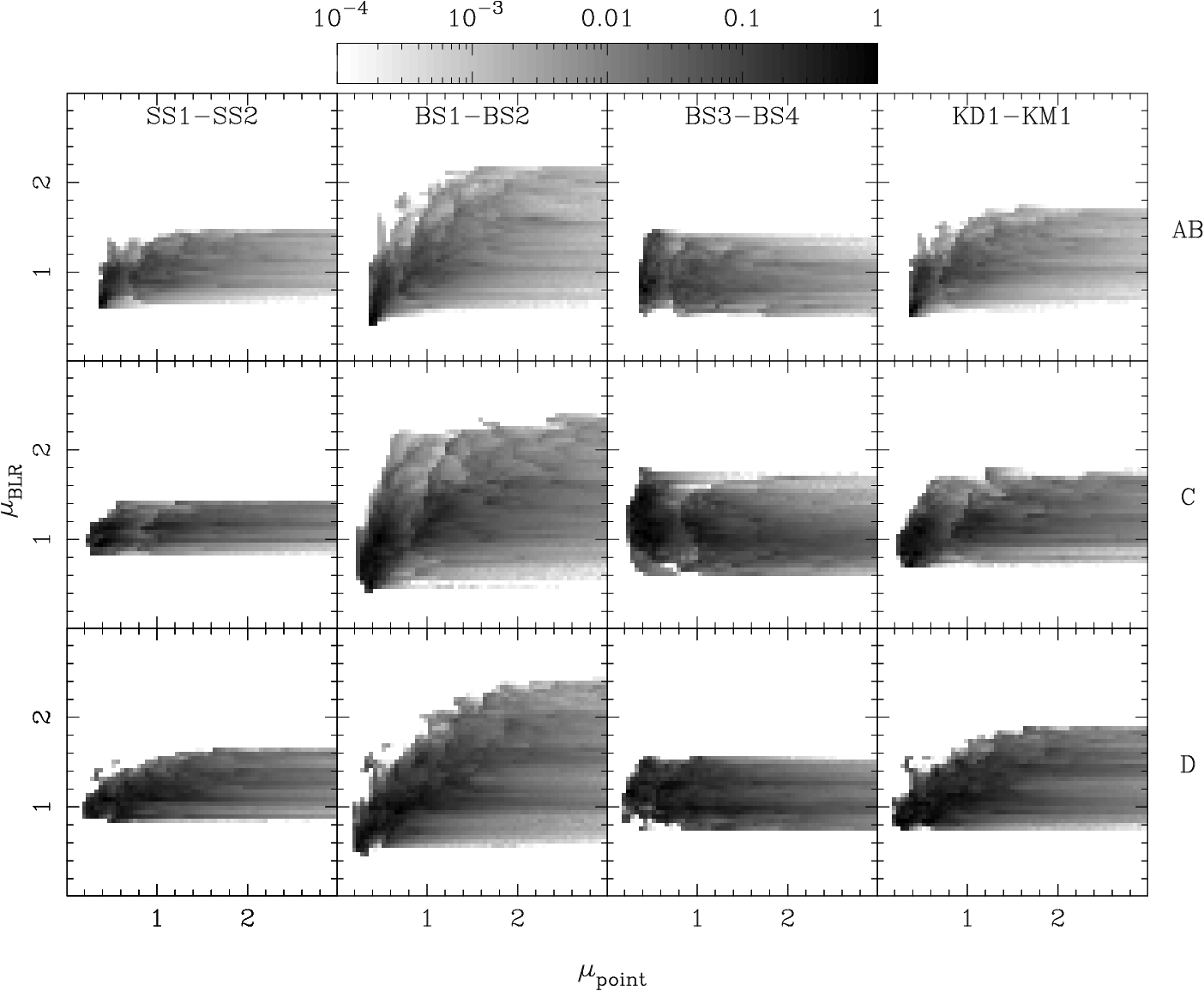,angle=270,width=5.8in}}
\caption[]{As for Figure~\ref{fig4}, except for the larger BLR models.
\label{fig5}}
\end{figure*}

Typically, gravitational microlensing  is observed via the photometric
monitoring         of          multiply         imaged         quasars
\citep{2000ApJ...529...88W,2002ApJ...572..729A}  with  no  program  of
spectroscopic monitoring of any system. To observe microlensing of the
BLR, therefore, it is  important to know whether expected fluctuations
are correlated with those of  the central continuum source; this would
be somewhat expected as the continuum source, which lies at the centre
of the BLR, `sees' similar caustic structure to the inner parts of the
BLR.   Figure~\ref{fig4} (smaller  BLR  source) and  Figure~\ref{fig5}
(larger BLR source) present the distributions of the magnifications of
the continuum source (taken to  be a single pixel in the magnification
map) versus the magnification  of the BLR for coincident observations.
Note  that  the  relative  probability,  displayed  in  grayscale,  is
presented logarithmically.

\begin{figure*}
\centerline{ \psfig{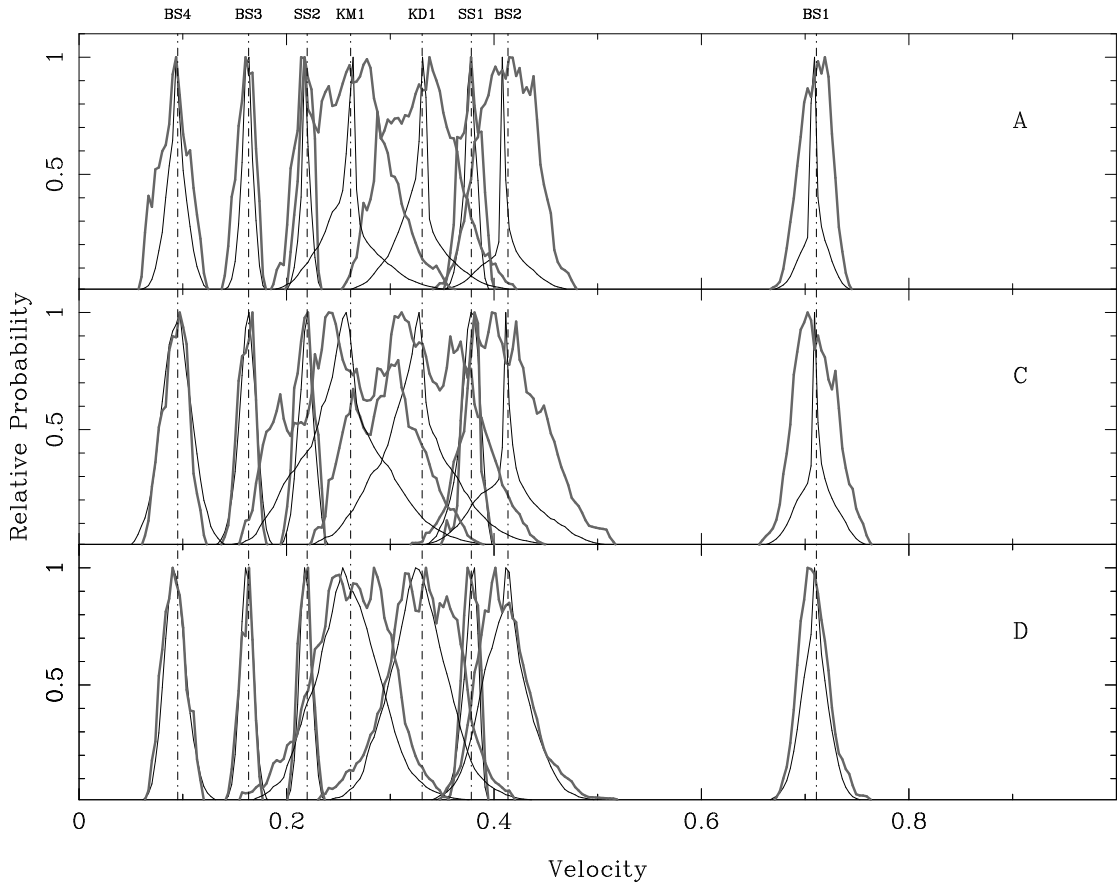}}
\caption[]{The  variation  in   the  centroid  velocity  for  positive
velocity  half of the  emission line  profile for  each of  the models
presented in this  paper. The solid black line is  for the smaller BLR
model, where as the thicker, grey line is for the corresponding larger
model.   The vertical dot-dashed  line indicates  the location  of the
unlensed centroid position, with the label for each model given at the
top of the line.\label{fig6}}
\end{figure*}

In examining Figures~\ref{fig4} and  \ref{fig5} it is clear that there
is small  scale structure  in the combined  probability distributions;
this is  due to  the finite size  of the magnification  maps employed,
with the structure due to the characteristics of the caustic networks.
Overall  trends, however, are  apparent. Firstly,  there is  a general
correlation between  the magnification of  the continuum and  the BLR.
For the  smaller BLR  source, this most  apparent for  the $BS1$-$BS2$
models.  As discussed previously, these models present a quite compact
surface  brightness  distribution  to  the  caustic  network  and  is,
therefore, more similar in scale to the continuum source.  In general,
the magnifications are correlated for  $\mu \leq 1$; this results from
the fact  that regions  of demagnification can  be extended  on scales
greater than an ER and hence  both the BLR and continuum region can be
demagnified together.   However, the correlation  of magnifications is
weaker  at $\mu  > 1$,  showing considerable  scatter for  all  of the
models.   This can be  understood in  terms of  the clustering  of the
caustic  network  in  regions  of high  magnification  which  exhibits
structure on quite small scales.   With this, the magnification of the
continuum mirrors this small  scale caustic structure, whereas the BLR
is  magnified  by a  weighted  average  of  the larger  scale  caustic
structure.   Interestingly,  for  all  the  models, there  is  also  a
non-zero probability that while the continuum source is being strongly
magnified, the over all BLR region is undergoing demagnification.  The
reverse of  this, however, appears  to be significantly rarer  in most
models.

The   situation  is   very   similar  for   the   larger  BLR   models
(Figure~\ref{fig5}).    As   expected   from  Figure~\ref{fig2},   the
distributions  of BLR  magnification  are somewhat  narrower than  the
smaller BLR model.   Again, no clear correlation of  the continuum and
BLR  magnification is  apparent with  the continuum  source undergoing
significant magnification while the BLR is relatively unmagnified.  It
is also apparent  that while there is the  general correlation between
the magnification  of the  two regions, there  is still  a significant
range over  which the continuum source can  be substantially magnified
while  the  BLR  undergoes  a  magnification  of  $\sim1$.   Hence,  a
microlensing fluctuation observed  in broadband photometric monitoring
will not  necessarily be  an indicator of  strong microlensing  of the
BLR.   Rather   than  spectroscopic  monitoring,   however,  broadband
monitoring  could be  combined  with observations  obtained through  a
narrow band filter  which covers a broad line  in the quasar spectrum.
With this, a plot similar to those presented in Figures~\ref{fig4} and
\ref{fig5} could be constructed and compared to simulations.

\subsection{Velocity Shifts}\label{vshifts}
As  well as  the  total magnification  of  the emission  lines, it  is
important  to  characterize  the  modification of  the  emission  line
profile due to differential magnification effects. This can be seen as
a shift in the velocity  centroid of the emission line. In undertaking
this,  however, it  is  important to  note  the $SS$  and $BS$  models
display surface brightness structure which is symmetrical in velocity,
and any gravitational lensing  magnification results in identical line
profile modification at positive and negative velocities, leading to a
centroid shift  of zero.  Therefore,  in the following study  only the
positive velocity  component of the emission lines  are considered for
the $SS$ and $BS$ models,  whereas the positive velocity and the total
emission line profile are considered for the $KD1$ and $KM1$ models.

Figure~\ref{fig6} presents  the distribution of  the measured centroid
of the positive  velocity component of the BLR  emission line for each
of the  models presented in this  paper.  The black  line presents the
distributions  for  the smaller  BLR  models  for  each lensed  image,
whereas the thicker,  grey line represents the larger  BLR models. The
vertical  dot-dashed  line   running  vertically  through  the  panels
represents     the      unlensed     location     of      the     line
centroid. Table~\ref{table2} summarizes the distributions presented in
Figure~\ref{fig6}, presenting their root  mean square (RMS); note that
in  these  values  are  percentages  of the  line  width  at  positive
velocities. Several features are apparent.  Firstly, the widths of the
distributions  are relatively insensitive  to the  microlensing model,
with  each   presenting  very  similar  forms   of  the  distribution.
Furthermore, the centroid shifts for the larger BLR models are similar
or are broader  than those for the smaller BLR  models. This result is
somewhat  surprising, given that  the larger  BLR models  undergo less
microlensing magnification, a point returned to below.

Table~\ref{table2}  summarizes  the  RMS  width of  the  distributions
presented  in  Figure~\ref{fig6}  as  a  percentage  of  the  positive
velocity line width.  The width of the distribution does depend on the
BLR model under  consideration, with the disk models  $KD1$ and $KM1$,
as   well   as  $BS3$   displaying   significantly  broader   centroid
distributions than the other models.  Examining the surface brightness
distributions presented  in Figure~\ref{fig1} it is  apparent that the
two disk models present  quite different surface brightness structures
in each of the velocity slices.  Hence these undergo differing degrees
of lensing  magnification, but  as their brightness  in each  slice is
non-negligible, this  can lead  to appreciable centroid  shifts. Other
models, such as $BS3$ and $BS4$ are luminous only in a narrow range of
velocity,  and magnification  of this  leads to  only  slight centroid
shifts.

\begin{figure*}
\centerline{ \psfig{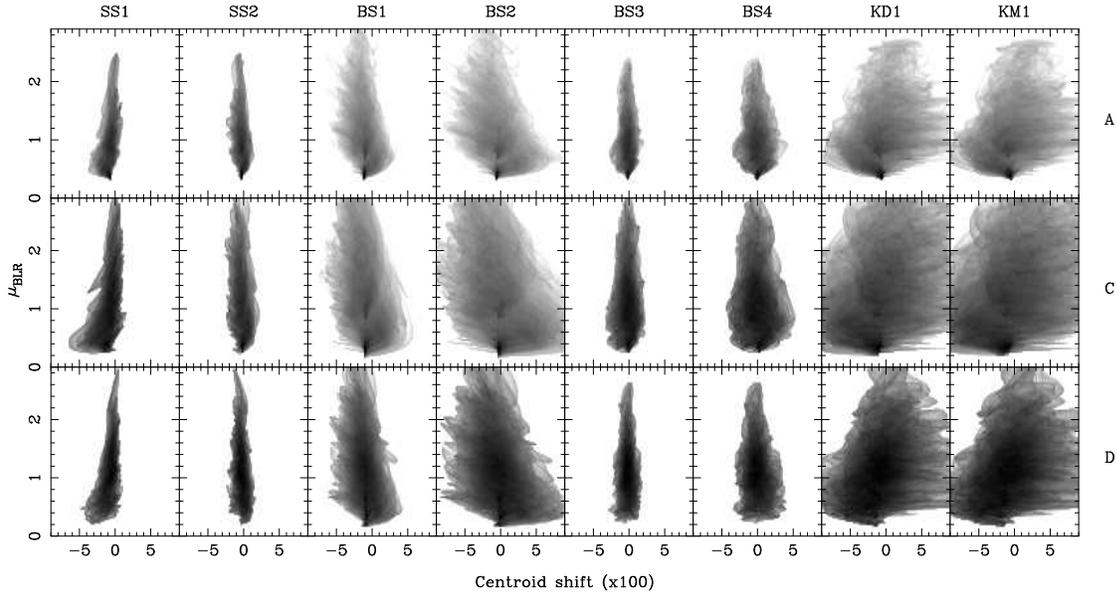}}
\caption[]{The   centroid   shift  in   velocity   versus  the   total
magnification  of the BLR  for each  of the  models presented  in this
paper.   This figure  considers solely  the smaller  BLR  models.  The
greyscale is identical to that presented in Figure~\ref{fig4}.
\label{fig7}}
\end{figure*}

\begin{figure*}
\centerline{ \psfig{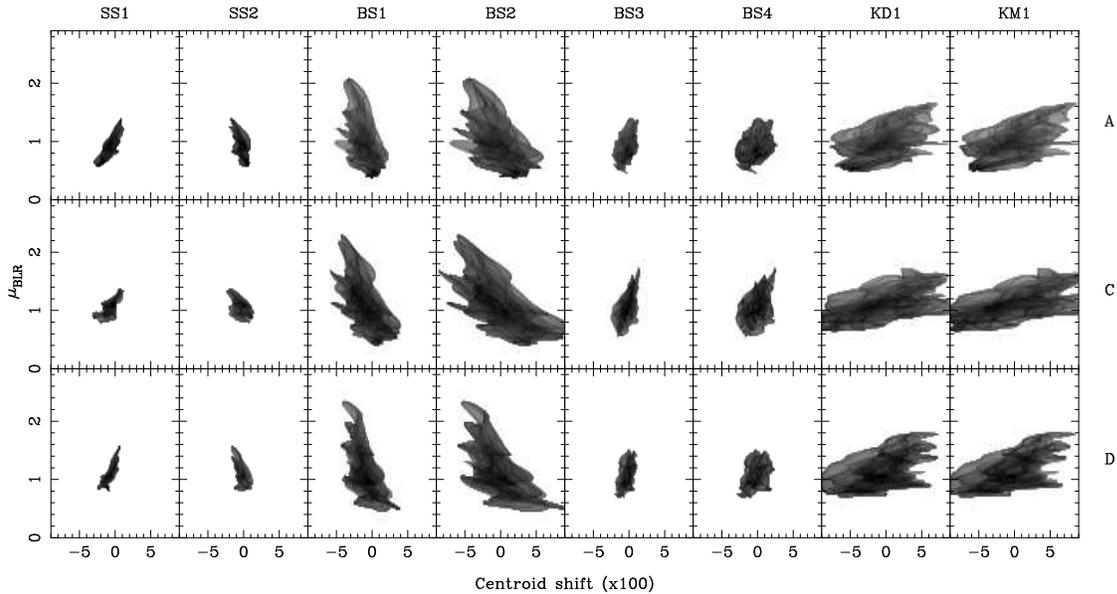}}
\caption[]{As for Figure~\ref{fig7}, except for the larger BLR models.
\label{fig8}}
\end{figure*}

Figures~\ref{fig7} \&~\ref{fig8} present the probability distributions
of the  centroid shift (presented in Figure~\ref{fig6})  as a function
of the total magnification of  the smaller BLR, with the corresponding
distributions for the larger BLR presented in Figure~\ref{fig8}. These
distributions are presented logarithmically on the same scale as those
in Figures~\ref{fig4} and \ref{fig5}. Again,  it is clear that some of
the structure in these distributions is due to the limited size of the
magnification maps, but some features are present. Firstly, it appears
that  the largest  centroid shifts  occur typically  at  lower overall
magnification. In this regime, only  a small range in velocity must be
magnified, leading to a centroid  shift but no substantial increase in
the  total line  flux.  For  the  total line  flux to  be enhanced,  a
substantial proportion  of all  velocity structure must  be magnified;
with  this,   the  centroid  shift  would  be   small.   Further,  the
distribution of the centroid shifts  appears, in a number of cases, to
be quite asymmetric.  This again  is related to the surface brightness
distribution as a function of velocity.

\begin{figure}
\centerline{ \psfig{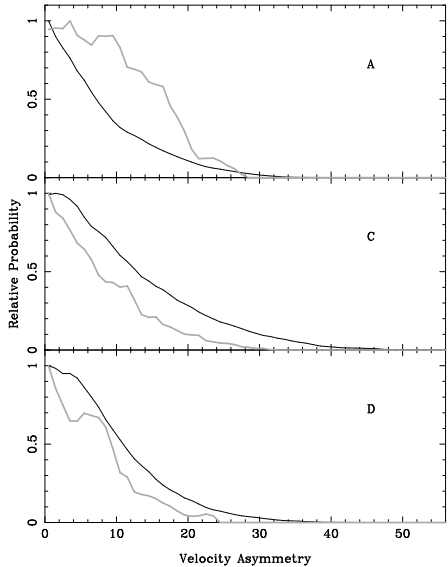}}
\caption[]{The asymmetry  in the  BLR emission line  for the  two disk
models presented in  this paper. The black line  denotes the asymmetry
for  the smaller  BAL radius  models,  whereas the  thicker grey  line
represents the larger models.
\label{fig9}}
\end{figure}

For the asymmetric surface brightness distributions ($KD1$ \& $KM1$) 
the line asymmetry was defined to be
\begin{equation}
A = 100 \times abs \left( \frac{ f_+ - f_- }{ f_+ - f_- }\right),
\end{equation}
where  $f_+$ is  the flux  in the  positive velocity  fraction  of the
emission line and  $f_-$ is the flux in  the negative velocity region.
Noting  that  the $KD1$  and  $KM1$  models  are identical  in  either
integrated positive  or negative velocity,  only one of these  need be
considered   further.   Figure~\ref{fig9}  presents   the  probability
distribution of  the line asymmetry  for the disk models  presented in
this paper,  with each horizontal  panel representing the  results for
various microlensing models. In each  panel, the black line is for the
smaller BLR models,  whereas the thicker, grey line  is for the larger
BLR  models.   The  distributions  for  the  various  models  are  not
dissimilar, showing  that substantial asymmetries in line  flux (up to
$\sim$20-30\%) can  result for  the smaller BLR  models when  they are
microlensed.   Surprisingly,  however,  is  the fact  that  while  the
magnification of  the larger BLR  models is typically smaller  for the
more  extended  BLR models,  the  substantial  line asymmetries  still
result.     This    regime   mirrors    that    earlier   probed    by
\citet{1990A&A...237...42S} who found that,  for the larger BLR models
they  were  considering, the  total  magnification  may  be mild,  but
substantial asymmetric  modification of the emission  line profile can
result.

\begin{table}
\centering
\begin{minipage}{70mm}
\caption{The widths of the centroid shift distributions
presented in Figure~\ref{fig6}\label{table2}}
\begin{tabular}{lcc} \hline \hline
Name & Small BLR (A,C,D) & Large BLR (A,C,D) \\ \hline
SS1 & (0.67,1.08,0.80)   &   (0.96,0.86,0.72) \\
SS2 & (0.52,0.75,0.58)   &   (0.65,0.85,0.65) \\
BS1 & (1.13,1.66,1.34)   &   (1.38,1,94,1.48) \\
BS2 & (2.05,3.06,2.43)   &   (2.49,3.51,2.67) \\
BS3 & (0.58,0.86,0.64)   &   (0.82,0.78,0.65) \\ 
BS4 & (1.00,1.50,1.11)   &   (1.40,1.18,1.09) \\
KD1 & (2.56,3.78,2.98)   &   (3.24,4.67,3.32) \\
KM1 & (2.79,4.10,3.25)   &   (3.51,5.00,3.57) \\
\hline
\end{tabular}
\end{minipage}
\end{table}

\section{Conclusions}\label{conclusions}
Recent studies  have reappraised the  scales of structure  in quasars,
with the indication  that the BLR, responsible for  the broad emission
lines  seen in  quasar spectra,  is smaller  than  previously thought.
This  reduction  in  size  makes  the region  more  sensitive  to  the
influence of gravitational microlensing.  This paper has examined this
influence  on eight models  of the  BLR, considering  the microlensing
parameters  of  the   multiply  imaged  quasar  Q2237+0305,  extending
previous studies into the high optical depth regime.

For the purpose of this study,  two source sizes were adopted; a small
source with a  radius of 1 ER and  a larger source with a  radius of 4
ER.  It was  found that  the  smaller source  can undergo  significant
magnification, with the total line  flux being enhanced by a factor of
2  on occasions.  At  the other  extreme, this  smaller source  can be
substantially  demagnified (with  respect to  the  mean magnification)
such that  its total  flux is  reduced to 20\%  of the  mean magnified
value. As  expected, the variations  are less dramatic for  the larger
sources,   with  a   typical  demagnification   to   $\sim80\%$,  with
magnification  extremes of $1.5\times\rightarrow2\times$.   In earlier
studies which  considered larger BLR  models, the total line  flux was
found to remain relatively unchanged during microlensing. If, however,
the revised  BLR sizes are  correct, this paper has  demonstrated that
substantial fluctuations  in the total  line flux should  result. This
has an  important consequence for studies of  gravitational lensing as
it implies that  the relative broad line flux between  images is not a
measure  of   the  relative  image   macromagnifications,  a  quantity
important to gravitational lensing modelling.

Furthermore, this paper investigated the degree of modification of the
form of the  BLR emission line via a measurement  of its centroid. The
majority  of models  considered possess  symmetric  surface brightness
structure in velocity space, and  the overall velocity centroid of the
emission  line   remains  unchanged  during   microlensing.   However,
considering  only one  half of  the emission  line it  was  found that
substantial   modification   of   the   emission  line   profile   can
result. Additionally, considering  disk models that present asymmetric
surface brightness  structure as  a function of  velocity, it  is seen
that  substantial  centroid shifts  of  the  entire  emission line  of
$\sim20\%$ can result.

Important differences were found  in the microlensed behaviour for the
models  under  consideration  in   this  paper,  with  the  degree  of
magnification and shift in the  line centroid being dependent upon the
surface brightness distribution as  a function of velocity. An obvious
example of this  would be the detection of  asymmetric modification of
the  overall emission  line profile,  indicating a  surface brightness
distribution  which is  asymmetric in  velocity, such  as  a structure
possessing rotation.   While it goes beyond this  current paper, these
results  reveal the possibility  of undertaking  detailed microlensing
tomography of the BLR  via spectroscopic monitoring of multiply imaged
quasars. However, current  microlensing monitoring programs focus upon
obtaining  broadband  photometry,   effectively  determining  the  the
microlensing  light   curves  for  the   continuum  source.   Previous
spectroscopic  studies  have  revealed  interesting BLR  profile  line
differences             between             various             images
\citep[e.g.][]{1989ApJ...338L..49F},     although     no    systematic
spectroscopic   program  has   been  undertaken   with   most  studies
consisting    of     single    or    double     epoch    observations
\citep[e.g.][]{1998MNRAS.295..573L}.     To   fully    determine   the
observational  implications of  this  study, an  investigation of  the
temporal  relationship between  the microlensing  of the  BLR  and the
continuum emitting source is  required, allowing the development of an
optimum override strategy such  that spectroscopic observations can be
obtained.  This is the subject of a forthcoming article.

\section{Acknowledgments}
Joachim Wambsganss is thanked for providing a copy of his microlensing
raytracing code  which was employed in  this study and  Scott Croom is
thanked for  helping unraveling what  bright means when  talking about
quasars.  The  authors  are  extremely  grateful  to  The  Centre  for
Astrophysics and Supercomputing  at Swinburne University of Technology
for  providing substantial computational  resources available  to this
project, and apologize to B. Conn,  J.  Chapman and I.  Klamer for the
mondas's screeching alarm when its cpus overheated.

\newcommand{\aap}{A\&A}
\newcommand{\apj}{ApJ}
\newcommand{\apjl}{ApJ}
\newcommand{\aaps}{AAPS}
\newcommand{\aj}{AJ}
\newcommand{\mnras}{MNRAS}

\end{document}